\documentclass{aa}
\usepackage{graphics,amssymb}
\begin{document}

\thesaurus{03(02.05.1; 04.19.1; 13.07.1)}

\title{Search for a possible space-time correlation between high energy
neutrinos and gamma-ray bursts}

\author{Montaruli T.\inst{1} 
\and Ronga F.\inst{2} for the MACRO collaboration$^{(*)}$}

\offprints{T. Montaruli}

\institute{INFN, Dipartimento di Fisica, Universit\`a di Bari, 
via Amendola 173, I-70126, Bari, Italy 
\and INFN, Laboratori Nazionali di Frascati, P.O. Box 13 I-00044,
Frascati, Italy
}

\date{Received \today / Accepted }

\titlerunning{Space-time correlation between $\nu$s and GRBs}
\authorrunning{Montaruli T. \& Ronga F.}

\maketitle

\begin{abstract}
We look for space-time correlations between 2233 gamma-bursts in the
Batse Catalogs and 894 upward-going muons 
produced by neutrino interactions in the rock below or inside MACRO. 
Considering a search cone of $10^{\circ}$ around GRB directions
and a time window of $\pm 200$ s we find 0 events to be compared to 
0.035 expected background events due to atmospheric neutrinos.
The corresponding upper limit (90$\%$ c.l.) is 
$0.87 \times 10^{-9}$ cm$^{-2}$ upward-going muons per
average burst.%
\end{abstract}
\section{Neutrino astronomy with MACRO}
Besides high energy gamma ray production resulting from 
$\pi^{0}$ decay, astrophysical beam dump models predict neutrino emission
from $\pi^{\pm}$ decay. Mesons are produced by accelerated protons interacting
with matter or photons in an accretion disk (Gaisser \cite{Gaisser}).  
The discovery of TeV gamma-ray emissions has enhanced the potential 
possibilities of this mechanism and the possible existence of such sources, 
but energies are not high enough to esclude syncroton radiation or
bremsstrahlung and inverse Compton production mechanisms. 
Neglecting photon absorbtion, it is expected that 
neutrino fluxes are almost equal to gamma ray ones and that the spectrum
has the typical form due to the Fermi 
acceleration mechanism: $\frac{dN}{dE} \propto E^{-2.0 \div 2.2}$.
Neutrinos produced in atmospheric cascades are background to the
search for astrophysical neutrinos for energies $\lesssim 10$ TeV.     
In fact, atmospheric neutrinos have a softer spectrum
than astrophysical neutrinos, since at energies $\gtrsim 100$ GeV
the decay length of mesons in the atmosphere becomes longer than 
atmospheric depth and the spectrum steepens (differential spectral index 
$\gamma\simeq$ 3.7). 
GRBs are possible sources of high energy $\nu$s: in the fireball scenario
the beam dump mechanism can lead to $\nu$ emission
(Halzen \cite{Halzen}, Waxman \cite{Bahcall}). 
The trancience of the GRB emission, even though
constrains the maximum neutrino energy to $\lesssim 10^{19}$ eV, improves
GRB association with measured neutrino events in underground detectors
using arrival directions and times, even though 
expected fluxes are much lower than atmospheric $\nu$ background.

The MACRO detector, located in the Hall B of the Gran Sasso underground 
laboratories, with a surface of $76.6 \times 12$ m$^{2}$ and a height of 9 m,
can indirectly detect neutrinos using a system of $\sim 600$ ton of
liquid scintillator to measure the time of flight of particles 
(resolution $\sim 500$ ps) and 
$\sim 20000$ m$^{2}$ of streamer tubes for tracking
(angular resolution better than 1$^{\circ}$ and pointing accuracy
checked using moon shadow detection (Ambrosio \cite{moon})). The time of flight
technique allows the discrimination between downward-going atmospheric muons
and upward-going events produced in the rock below 
(average atmospheric neutrino energy $\langle E_{\nu} \rangle \sim 100$ GeV)
and inside ($\langle E_{\nu} \rangle \sim 4$ GeV) 
the detector by neutrinos which have crossed the Earth.
Between $\sim 31 \times 10^{6}$ atmospheric muons, a sample of 
909 upward-going muons is selected with an automated analysis. 
The data taking has begun since 
March 1989 with the incomplete detector (Ahlen \cite{macro95}) 
and since April 1994 with the full detector (Ambrosio \cite{macro98}).
In our convention $1/\beta = \Delta T c/L$, calculated from the measured 
time of flight $\Delta t$ and the track length between the scintillator 
layers, is $\sim 1$ for downward-going muons and $\sim -1$ for 
upward-going muons. 
Events with $-1.25 < 1/\beta <-0.75$ are selected.

We look for a statistically significant excess of
$\nu$ events in the direction of known $\gamma$ and $X$-ray sources
(a list of 40 selected sources, 129 sources of the $2^{nd}$ Egret Catalogue,
2233 Batse GRB, 220 SN remnants, 
7 sources with $\gamma$ emission above 1 TeV)
with respect to fluctuations of the atmospheric $\nu$ background.
The expected background from atmospheric $\nu$s is calculated in 
declination bands of $\pm 5^{\circ}$ around the declination of the source
mixing for 100 times local coordinates and times of upward-going $\mu$s. 
We calculate flux limits in half-cones of $3^{\circ}$
taking into account reduction factors due to the signal fraction
lost outside the cone (which depends on $\nu$ spectrum, kinematics
of CC interaction, $\mu$ propagation in the rock, MACRO angular resolution).
We do not find any signal evidence from known sources or of clustering of
events (we measure 
89 clusters of $\ge 3$ events and expect 
81.2 of them in a cone of $3^{\circ}$). 
Muon flux limits for some sources are:
2.5 for Crab Nebula, 5.6 for MRK421, 3.71 for Her X-1, 0.45
for Vela Pulsar, 
0.65 for SN1006 in units of 
$\times 10^{-14}$ cm$^{-2}$ s$^{-1}$. For most of the
considered sources MACRO gives the best flux limits compared
to other underground experiments.  
\section{Space-time correlations between GRBs and upward-going muons}
\begin{figure}
\resizebox{8.7cm}{6.cm}{\includegraphics{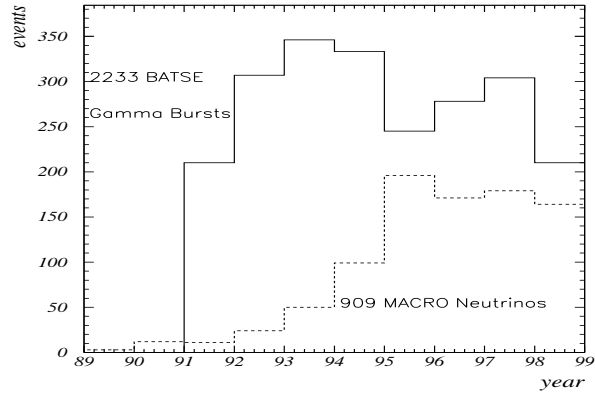}}
\caption{MACRO upward-going muon (dashed line) and Batse GRB (solid line)
distributions versus year.}
\label{fig1}
\end{figure}

We look for correlations with 2233 GRBs in the
Batse Catalogs 3B and 4B (Meegan \cite{Meegan}) 
collected since 21 Apr. 1991 to 5 Oct. 1998 and 894 of the 909 
upward-going muons detected by MACRO during this period
(see Fig.~\ref{fig1}). 
Considering Batse angular accuracy, we estimate
that a half-cone of $20^{\circ}$ ($10^{\circ}$) contains 99.8$\%$ (96.8$\%$)
of $\nu$ sources (if GRB sources are $\nu$ sources, too). 
The same percentages of upward-going muons are contained in these  
half-cones from the GRB sources. As a matter of fact, we calculate 
via Monte Carlo the fraction of signal lost, which depends on the 
$\nu$ spectral index, multiple scattering of muons during propagation
in the rock and MACRO angular resolution. Using various cone apertures, 
we estimate that this fraction is negligible for $\ge 10^{\circ}$. 
 
\begin{figure}
\resizebox{\hsize}{!}{\includegraphics{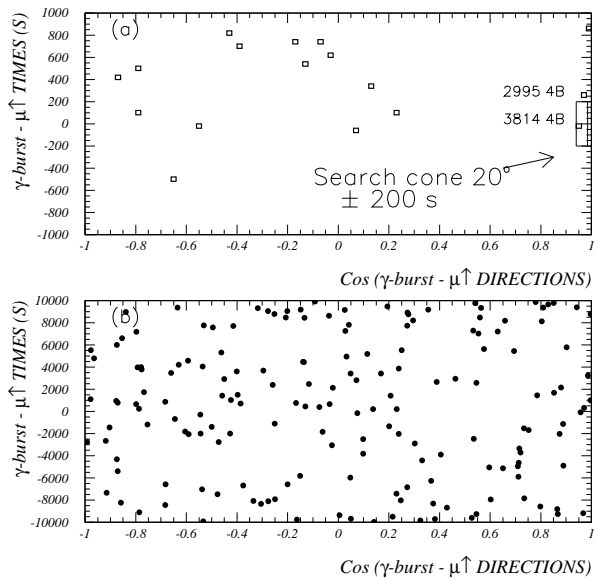}}
\caption{Difference in detection times vs cosine of angular separation 
between Batse GRBs and upward-going $\mu$s. (a) and (b) have different time 
scales. In (a) the $\pm 200$ s - $10^{\circ}$,$20^{\circ}$ windows are 
indicated.}
\label{fig2}
\end{figure}

The area for upgoing muon detection in the direction of the GRBs
averaged over all the bursts is 119 m$^{2}$. Its value
is small because MACRO is sensitive to neutrinos only in the lower hemisphere
and because it was incomplete in the period 1991-1994. 
We find no statistically significant correlation between neutrino event
and GRB directions and detection times.
As shown in Fig.~\ref{fig2}, we find no events in a window 
of $\pm 200$ inside $10^{\circ}$ from GRB directions and 1 event inside 
$20^{\circ}$, which was measured after 39.4 s from the Batse GRB of 
22 Sep. 1995 (4B 950922). 
For this burst the radius of the positional
error box in the Batse catalog is 3.86$^{\circ}$, much smaller than 
the angular distance of 17.6$^{\circ}$ at which we find the 
neutrino event.
The expected number of atmospheric $\nu$ events is computed 
with the delayed coincidence technique. We expect 0.035 (0.075)
events in $10^{\circ}$ ($20^{\circ}$).
The corresponding upper limits (90$\%$ c.l.) for the upward-going muon flux
are $0.87 \times 10^{-9}$ cm$^{-2}$ ($1.44 \times 10^{-9}$ cm$^{-2}$) 
upward-going muons per average burst. 
These limits exclude an extreme cosmic string-type model
reported in (Halzen \cite{Halzen}), which results in $10^{-1} \, 
\mu$ cm$^{-2}$, while according to a fireball scenario model in 
(Waxman \cite{Bahcall}), 
a burst at a distance of 100 Mpc producing 
$0.4 \times 10^{51}$ erg in neutrinos of around $10^{14}$ eV
would produce $\sim 6 \times 10^{-11}$ cm$^{-2}$ upward-going muons.

(*)
{\bf The MACRO Collaboration}\\
\nobreak
\begin{center}
\pretolerance=1000
M.~Ambrosio$^{12}$, 
R.~Antolini$^{7}$, 
C.~Aramo$^{7,n}$,
G.~Auriemma$^{14,a}$, 
A.~Baldini$^{13}$, 
G.~C.~Barbarino$^{12}$, 
B.~C.~Barish$^{4}$, 
G.~Battistoni$^{6,b}$, 
R.~Bellotti$^{1}$, 
C.~Bemporad $^{13}$
% ...
\\
\end{center}
\footnotesize
1. Dipartimento di Fisica dell'Universit\`a di Bari and INFN, 70126 
Bari,  Italy \\
2. Dipartimento di Fisica dell'Universit\`a di Bologna and INFN, 
 40126 Bologna, Italy \\
3. Physics Department, Boston University, Boston, MA 02215, 
USA \\
4. California Institute of Technology, Pasadena, CA 91125, 
USA \\
5. Department of Physics, Drexel University, Philadelphia, 
PA 19104, USA \\
6. Laboratori Nazionali di Frascati dell'INFN, 00044 Frascati (Roma), 
Italy \\
7. Laboratori Nazionali del Gran Sasso dell'INFN, 67010 Assergi 
(L'Aquila),  Italy \\
8. Depts. of Physics and of Astronomy, Indiana University, 
Bloomington, IN 47405, USA \\
9. Dipartimento di Fisica dell'Universit\`a dell'Aquila  and INFN, 
 67100 L'Aquila,  Italy \\
10. Dipartimento di Fisica dell'Universit\`a di Lecce and INFN, 
 73100 Lecce,  Italy \\
11. Department of Physics, University of Michigan, Ann Arbor, 
MI 48109, USA \\	
12. Dipartimento di Fisica dell'Universit\`a di Napoli and INFN, 
 80125 Napoli,  Italy \\	
13. Dipartimento di Fisica dell'Universit\`a di Pisa and INFN, 
56010 Pisa,  Italy \\	
14. Dipartimento di Fisica dell'Universit\`a di Roma ``La Sapienza" and INFN, 
 00185 Roma,   Italy \\ 	
15. Physics Department, Texas A\&M University, College Station, 
TX 77843, USA \\	
16. Dipartimento di Fisica Sperimentale dell'Universit\`a di Torino and INFN,
 10125 Torino,  Italy \\	
17. L.P.T.P., Faculty of Sciences, University Mohamed I, B.P. 524 Oujda, Morocco \\
$a$ Also Universit\`a della Basilicata, 85100 Potenza,  Italy \\
$b$ Also INFN Milano, 20133 Milano, Italy\\
$c$ Also Istituto TESRE/CNR, 40129 Bologna, Italy \\
$d$ Also Scuola Normale Superiore di Pisa, 56010 Pisa, Italy\\
$e$ Also Dipartimento di Fisica, Universit\`a di Roma Tre, Roma, Italy \\
$f$ Also Universit\`a di Trieste and INFN, 34100 Trieste, 
Italy \\
$g$ Also Dipartimento di Energetica, Universit\`a di Roma, 
 00185 Roma,  Italy \\
$h$ Also Institute for Nuclear Research, Russian Academy
of Science, 117312 Moscow, Russia \\
$i$ Also Institute for Space Sciences, 76900 Bucharest, Romania \\
$l$ Swarthmore College, Swarthmore, PA 19081, USA\\
$m$ RPD, PINSTECH, P.O. Nilore, Islamabad, Pakistan \\
$n$ Also INFN Catania, 95129 Catania, Italy\\
$o$ Also Department of Physics, Pennsylvania State University, 
University Park, PA 16801, USA
\end{document}